\documentclass[twocolumn,showpacs,preprintnumbers,amsmath,amssymb]{revtex4}
\usepackage{graphicx}% Include figure files
\usepackage{dcolumn}% Align table columns on decimal point
\usepackage{bm}% bold math

\begin{document}

\preprint{APS/123-QED}

\title{Longitudinal SDW order in a quasi-1D Ising-like quantum antiferromagnet}% Force line breaks with \\

\author{S. Kimura$^1$, M. Matsuda$^2$, T. Masuda$^3$, S. Hondo$^3$, K. Kaneko$^4$, N. Metoki$^4$,\\ M. Hagiwara$^1$, T. Takeuchi$^5$, K. Okunishi$^6$, Z. He$^7$, K. Kindo$^7$, T. Taniyama$^8$ and M. Itoh$^8$}
% \altaffiliation[Also at ]{Physics Department, XYZ University.}%Lines break automatically or can be forced with \\
%\author{Second Author}
%\email{Second.Author@institution.edu}
\affiliation{
$^1$KYOKUGEN, Osaka University, Machikaneyama 1-3, Toyanaka 560-8531, Japan \\ 
$^2$Quantum Beam Science Directorate, Japan Atomic Energy Agency, Tokai, Ibaraki 319-1195, Japan\\ 
$^3$International Graduate School of Arts and Sciences, Yokohama City University, 22-2, Seto, Kanazawa-ku, Yokohama, Kanagawa, 236-0027, Japan\\ 
$^4$Advanced Science Research Center, Japan Atomic Energy Agency, Tokai, Ibaraki 319-1195, Japan\\ 
$^5$Low temperature Center, Osaka University, Machikaneyama 1-1, Toyonaka, Osaka, Japan\\ 
$^6$Department of physics, Niigata University, Niigata 950-2181, Japan\\ 
$^7$The Institute for Solid State Physics, University of Tokyo, Kashiwa, Chiba 277-8581, Japan\\ 
$^8$Material and Structure Laboratory, Tokyo Institute of Technology, 4259 Nagatsuka, Midori, Yokohama, 226-850, Japan
}
%\author{Charlie Author}
%\homepage{http://www.Second.institution.edu/~Charlie.Author}
%\affiliation{
Second institution and/or address\\
This line break forced% with \\

\date{\today}% It is always \today, today,
             %  but any date may be explicitly specified

\begin{abstract}
From neutron diffraction measurements on a quasi-1D Ising-like Co$^{\rm 2+}$ spin compound BaCo$_{\rm 2}$V$_{\rm 2}$O$_{\rm 8}$, we observed an appearance of a novel type of  incommensurate ordering in magnetic fields. This ordering is essentially different from the N{\' e}el-type ordering, which is expected for the classical system, and is caused by quantum fluctuation inherent in the quantum spin chain. A Tomonaga-Luttinger liquid (TLL) nature characteristic of the gapless quantum 1D system is responsible for the realization of the incommensurate ordering. 
\end{abstract}

% PACS, the Physics and Astronomy
                             % Classification Scheme.
%\keywords{Suggested keywords}%Use showkeys class option if keyword
                              %display desired
\maketitle
The quasi-one-dimensional (1D) antiferromagnets, in which spin chains are coupled only weakly by interchain interactions, involve significant quantum fluctuation and often display exotic behavior. As expected for the classical models, the antiferromagnet in three-dimension typically develops a usual N{\' e}el order with antiparallel alignments of neighboring spins when it is cooled. Owing to the strong quantum fluctuation, however, qualitatively different situations from the classical models can appear for the quasi-1D antiferromagnets. Several exotic states have been discovered in the quasi-1D antiferromagnets, such as the valence-bond-solid ground state~\cite{Affleck} or the spin-Peierls state~\cite{Pytte}. In this letter, we show that in the system with Ising-like anisotropy in magnetic fields, the symmetry breaking for the long range order takes place in a very unusual way. We have found a novel type of density-wave-like incommensurate ordering in a quasi-1D Ising-like Co$^{\rm 2+}$ spin compound BaCo$_{\rm 2}$V$_{\rm 2}$O$_{\rm 8}$.  This curious ordering originates from instability of the peculiar quantum critical nature, characterized by an incommensurate modulation wave vector in one-dimensional chain, that has no classical analog. 

The aniferromagnetic chain in magnetic fields applied along the longitudinal $z$-direction is described by the following $XXZ$ Hamiltonian: 
 \begin{eqnarray}
{\mathcal H} = J \sum_{\rm i} \left\{ S^{\rm z}_{\rm i}S^{\rm z}_{\rm i+1}+\epsilon (S^{\rm x}_{\rm i}S^{\rm x}_{\rm i+1}+S^{\rm y}_{\rm i}S^{\rm y}_{\rm i+1})\right\} \nonumber\\
- g{\mu}_{\rm B} \sum_{\rm i}{S_{\rm i,z}}H
\end{eqnarray}
where, $J$ ($>$0) is an antiferromagnetic exchange constant, $\epsilon$ an anisotropic parameter, $g$ a $g$-value, ${\mu}_{\rm B}$ the Bohr magneton and $H$ the magnetic field. The system with $\epsilon$  $<$ 1, $\epsilon$  = 1 and $\epsilon$  $>$ 1 corresponds to the Ising-like, Heisenberg and $XY$-like antiferromagnetic chain, respectively. The quantum fluctuation, stemming from non-commutative properties of spin operators, plays a crucial role in determining its ground state properties. Indeed, the exact quantum mechanical ground state of the Heisenberg chain with spin $S$ = 1/2, found by Bethe in 1931~\cite{Bethe}, is a spin liquid state with no order, showing that the quantum fluctuation in this case disturbs the system in taking a long range ordered state even at absolute zero Kelvin. After a theoretical finding that the spin chain with $S$ = 1/2 can be represented by a pseudo-fermion model~\cite{Leib,Bulaevskii}, an important conclusion has been extracted~\cite{Haldane,Luther}, i.e. the spin liquid state belongs to a universality class called a Tomonaga-Luttinger liquid (TLL). 
\begin{figure}
\includegraphics[width=8.5cm,clip]{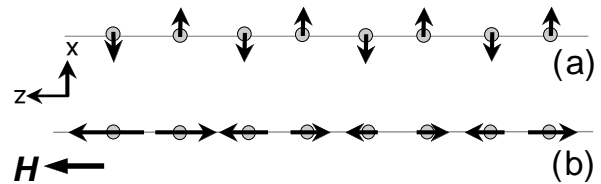}
\caption{Spin correlations in the TLL phase of the $S$ = 1/2 1D $XXZ$ antiferromagnet. (a) The staggered correlation of the transverse spin components, which is expressed as $<S^{\rm x}_{\rm 0}S^{\rm x}_{\rm r}> {\simeq} (-1)^{r}r^{-{\eta}_{\rm x}}$. (b) The incommensurate correlation of the longitudinal spin component, expressed as $<S^{\rm z}_{\rm 0}S^{\rm z}_{\rm r}>-m^2 {\simeq} {\rm cos}(2k_{\rm F}r)r^{-{\eta}_{\rm z}}$. $r$ is a distance between the spins. $\eta$$_{\rm x}$ and $\eta$$_{\rm z}$ are Tomonaga-Luttinger exponents, which satisfy a relation $\eta$$_{\rm x}$$\eta$$_{\rm z}$ = 1.}
\end{figure}
The TLL includes most 1D quantum systems having a gapless elementary excitation with a linear energy dispersion, such as a linear chain conducting electron system. In general, a TLL has no long range order even at zero Kelvin, but it is in a quantum critical state with intrachain correlations of algebraic decay~\cite{Luther}. A distinctive feature of the TLL in the spin chain is that two kinds of the correlation develop there as shown in Fig. 1~\cite{Haldane,Luther,Ishimura,Muller}. One is the staggered correlation of the transverse spin component perpendicular to the chain direction. The other is the incommensurate correlation of the longitudinal component parallel to the chain. One may associate the former with the usual N{\' e}el order, but the later has no classical analog and is peculiar to the quantum spin chain.  
\begin{figure}
\includegraphics[width=7cm,clip]{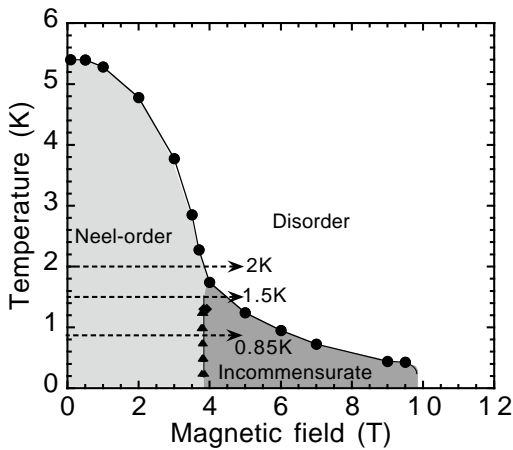}
\caption{Phase diagram of BaCo$_{\rm 2}$V$_{\rm 2}$O$_{\rm 8}$ in magnetic fields applied along the $c$-axis. Neutron scattering measurements are conducted in the region, shown by allows.}
\end{figure}

The TLL nature which appears in the $S$ = 1/2 Ising-like antiferromagnetic spin chain, brings the curious ordering found in this study. The Ising-like chain shows an interesting quantum phase transition in magnetic fields~\cite{Haldane,Yang}. Different from the Heisenberg system, the ground state of the Ising-like chain has a long range order at zero magnetic field, because the Ising-like anisotropy stabilizes the N{\' e}el state~\cite{Orbach}. The external field, however, induces the strong quantum fluctuation, that drives the system into the TLL phase above a certain critical field. The idea underlying our results concerns with the real quasi-1D compounds, in which interactions between the chains inevitably exist~\cite{Sakai,Suzuki,Okunishi,Kimura}. The interchain interactions make the TLL phase unstable, and can thereby lead the system into the long range ordering at a finite temperature. On this situation, between two kinds of the correlation mentioned above, the most dominant one will grow rapidly and build up the order. An important point for the Ising-like chain is that the incommensurate correlation is enhanced above the critical field, in contrast to the Heisenberg system, for which the staggered correlation is always dominant in a magnetic field~\cite{Bogoliubov}. Thus, we propose that a density-wave-like incommensurate ordering appears for the quasi 1D Ising-like antiferromagnet in the field-induced region, when the temperature is lowered enough to make the interchain interaction relevant~\cite{Suzuki,Okunishi,Kimura}. In this ordered state, the spins align to be collinear along the chain direction with modulation of those amplitude, characterized by the incommensurate wave number 2$k_{\rm F}$. This ordering is essentially different from that expected for the classical Ising-like antiferromagnet with no quantum fluctuation. The field-induced transition in the classical case is a spin-flop type, and the N{\' e}el order of the transverse component, which relates to the breaking of rotational symmetry around the field, appears in the field-induced region as well as the Heisenberg antiferromagnet~\cite{Fisher}. On the other hand, the incommensurate ordering discussed here occurs as a consequence of the breaking of quasi-continuous translation symmetry. According to the pseudo-fermion model, this incommensurate ordering is regarded as a charge density wave (CDW) ordering for the pseudo-fermion. The modulation wave number 2$k_{\rm F}$ of this incommensurate ordering can be easily tuned by applying magnetic fields, since it is determined by the magnetization per spin $m$ as 2$k_{\rm F}$ = ${\pi}$(1  - 2$m$)~\cite{Haldane}. This means that the periodicity of the incommensurate structure varies continuously with the field strength. This property is also different from that of the density-wave ordering in the conducting electron systems, of which 2$k_{\rm F}$ depends on the electron occupation number and is therefore little affected by external perturbations. The density-wave ordering in the conducting electron systems is arisen by the nesting of Fermi surface, whereas the incommensurate ordering in the quasi 1D Ising-like $XXZ$ case is entirely due to the quantum fluctuation inherent in the system. 

In order to find this curious ordering, neutron diffraction measurements, which provide direct information about the spin structure, are particularly useful. For this measurement, we adopt the Co$^{\rm 2+}$ spin system BaCo$_{\rm 2}$V$_{\rm 2}$O$_{\rm 8}$, in which magnetic Co-O chains are running along the crystallographic $c$-axis~\cite{Wichmann}. Recently, this compound was revealed to be a good realization of quasi-1D $S$ = 1/2 Ising-like antiferromagnet with the transition field $H_{\rm c}$ ${\simeq}$ 3.9 T [19, 20] that can be achieved in current neutron facilities. The field-temperature phase diagram  of BaCo$_{\rm 2}$V$_{\rm 2}$O$_{\rm 8}$, obtained from our thermodynamic measurements~\cite{Kimura,He1}, are depicted in Fig. 2. BaCo$_{\rm 2}$V$_{\rm 2}$O$_{\rm 8}$ undergoes N{\' e}el ordering at $T_{\rm N}$ = 5.4 K at zero magnetic field, but the ordered temperature is rapidly lowered by the external field along the chain, which corresponds the easy axis direction~\cite{He1}. The suppression of the N{\' e}el order by magnetic fields can be understood by an appearance of the TLL nature in the field, which was mentioned for the Ising-like chain before ~\cite{Kimura2}. However, we recently found that at very low temperatures below 1.8 K, another ordered phase emerges in the field-induced region above $H_{\rm c}$ ${\simeq}$ 3.9 T~\cite{Kimura}. In the ordered phase in the field-induced region, we expect a realization of the incommensurate spin structure.

We have performed the neutron diffraction measurements in the following condition. A single crystal of BaCo$_{\rm 2}$V$_{\rm 2}$O$_{\rm 8}$, having a shape of a plate with 6 ${\times}$ 6 ${\times}$ 26 mm$^{\rm 3}$ and a weight of  1.5 g, was used for neutron elastic scattering measurements. A space group of BaCo$_{\rm 2}$V$_{\rm 2}$O$_{\rm 8}$ is I41/$acd$. The neutron scattering measurements were carried out with the thermal neutron triple-axis spectrometer TAS-2 at the JRR-3 reactor at Japan Atomic Energy Agency. Incident neutrons are monochromatized by the (0 0 2) reflection of  
pyrolytic graphite (PG) crystal, and contamination from higher-order beams was effectively eliminated using a PG filter. The measurements were conducted in horizontal fields up to 5 T. A split-pair superconducting magnet manufactured by Oxford Instruments, UK, was used for the field generation. The fixed incident neutron energy was 13.7 meV. The horizontal collimator sequence was guide-80'-80'-open. The cooling of the sample was achieved by a $^{\rm 3}$He-$^{\rm 4}$He dilution refrigerator.
\begin{figure}
\includegraphics[width=7cm,clip]{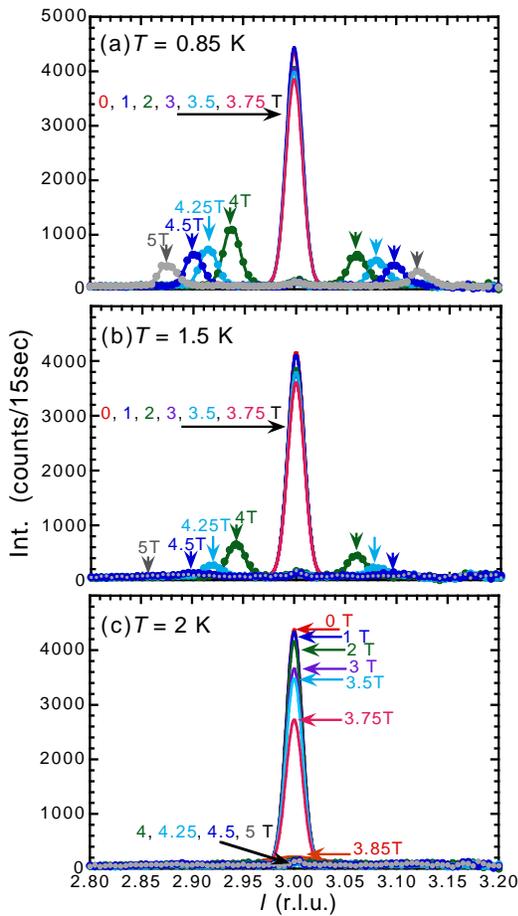}
\caption{Magnetic field dependence of neutron diffraction profiles of (4, 0, $l$) scan measured at temperature $T$ = 0.85 K (a), 1.5 K (b) and 2 K (c). Solid lines are the results of fits to Gaussian functions.}
\end{figure}

Now let us turn to our neutron diffraction investigation of the spin structure in BaCo$_{\rm 2}$V$_{\rm 2}$O$_{\rm 8}$. The measurement was conducted in magnetic fields up to 5 T along the $c$-axis and at temperature below 2 K. Figure 3 shows the field dependence of the scan profile of the (4 0 $l$) reflection in the fields.  In neutron scattering measurements, only the component of the magnetization that is perpendicular to the scattering vector contributes to the scattering intensity~\cite{Squires}. Thus, to detect the magnetic ordered component parallel to the chain direction, we adopt the $l$-scan around (4 0 3). In fact,
other reflections such as (4 0 1) are more suitable to detect the magnetic component along the $c$ axis. However, these reflections do not satisfy the configurational condition of our measurement system, which is restricted by the windows of the sprit-pair magnet. At zero magnetic field,  a peak at (4 0 3), which corresponds to the N{\' e}el order of the magnetic moments along the chain, is observed. According to the extinction rule, the (4 0 3) nuclear reflection in I41/$acd$ space group is prohibited. Thus, the observed (4 0 3) peak is purely magnetic, and we confirmed a disappearance of the peak above $T_{\rm N}$ = 5.4 K. The (4 0 3) peak gradually diminishes with increasing the field up to 3.75 T, and then a sudden change of the scan profile occurs around the transition field $H_{\rm c}$. In the field-induced region above $H_{\rm c}$, two peaks at positions incommensurate with the underlying lattice appear at temperatures below 1.5 K. The sudden change of the scan profile reflects the fact that the transition at $H_{\rm c}$ is weakly first order as suggested from our previous thermodynamic measurements~\cite{Kimura}. A tiny peak at (4 0 3) remains in the field-induced region, but its origin is not clear at the moment. The positions of the peaks are plotted in Fig. 3. The difference between the peak positions in field ascending and descending processes is small. 
\begin{figure}
\includegraphics[width=7cm,clip]{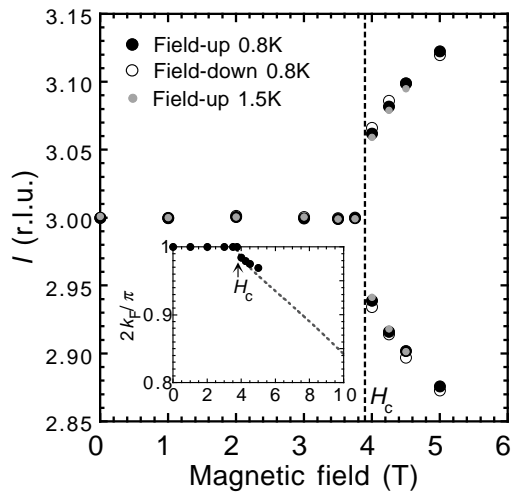}
\caption{Magnetic field dependence of the peak position of the observed neutron scan profile. The data are extracted from the least-square fits to the neutron scan profile. Inset shows the field dependence of a normalized incommensurate modulation 2$k_{\rm F}$/${\pi}$. The theoretical curve is obtained by the calculation based on the Bethe anzats exact theory.}
\end{figure}
The incommensurate peaks shift continuously in such a way that a distance between the two peaks increases with increasing the field. Slight temperature dependence of the peak positions is found for $H$ $>$ $H_{\rm c}$. The line width for all the peaks, observed in the ordered region, is within a resolution limit of our apparatus, and we confirm that the incommensurate peak, observed from the $h$-scan, is also resolution-limited. The resolution-limited peaks correspond to the development of the long range order. The scan profiles, observed at 4.5 T, show that the Bragg peaks change to broad defuse ones at 1.5 K, that is just above the ordering temperature, and then disappear with increasing the temperature. Our observation unambiguously demonstrates a realization of the incommensurate ordering in the field-induced phase of BaCo$_{\rm 2}$V$_{\rm 2}$O$_{\rm 8}$. In our experimental accuracy, no peak is found in the $l$-scan profile around (0 0 1) at which the neutron cross section includes solely transverse component of magnetic moment. The result indicates the density-wave-like ordering with collinear spin alignments along the chain directions in the field-induced  phase. Satellite reflections, coming from a higher harmonic Fourier components of the incommensurate modulation for the ordered structure, are not observed in the scan profile of the (4 0 $l$) reflection. Thus, the incommensurate modulation is suggested not to be square-wave like but is close to proper sinusoidal. In the inset of Fig. 4, we plot the field dependence of a normalized incommensurate modulation 2$k_{\rm F}/{\pi}$ along the $c$-axis, which is given by a relation 2$k_{\rm F}/{\pi}$ = 1 - ${\delta}$.  Taking into account the fact that four Co$^{\rm 2+}$ ions are included in a chemical unit of BaCo$_{\rm 2}$V$_{\rm 2}$O$_{\rm 8}$ along the chain, the ${\delta}$ is obtained from 1/8 of a distance between two incommensurate peaks. As anticipated, the 2$k_{\rm F}/{\pi}$ continuously decreases with increasing the field above $H_{\rm c}$. The experimental result slightly deviates from the theoretical prediction 2$k_{\rm F}$ = ${\pi}$(1- 2$m$)~\cite{Haldane} with increasing field as shown in the inset of Fig. 4. The theoretical curve is obtained by the calculation based on the Bethe ansatz exact theory with the parameters $J/k_{\rm B}$ = 65 K, $\epsilon$ = 0.46 and $g$ = 6.2, which are estimated from the magnetization curve ~\cite{Kimura2}. The observed two incommensurate peaks reflect an existence of two kinds of domain for the ordered structure. The left and right peaks correspond to the domain with the modulation wave vector pointing parallel and antiparalell to the field direction in the chain, respectively. The difference of the intensity between the two peaks probably originates from the imbalance in two domains caused by the external field. 

The authors are grateful to Dr. T. Suzuki for valuable discussions. This work was partly supported by Grants-in-Aid for Young Scientists (B) (No. 18740183, No. 1870230 and No. 19740215), for Scientific Research on Priority Areas "High Field Spin Science in 100 T" (No. 451) and "Novel State of Matter Induced by Frustration" (No. 19052004 and No. 19052008) and for Scientific Research (B) (No. 20340089) from the Ministry of Education, Science, Sports, Culture and Technology (MEXT) of Japan.

%%%%%%%%%%%%%%%%%%%%%%%%%%%%%%%%%%%%%%%%%%%%%%%%%%%%%%%%%%%%%%
%%%%%%%%%%%%

\end{document}